\documentclass[12pt]{elsart}
\usepackage{amsmath}
\begin{document}
\begin{frontmatter}
\title{Trilinearization and Localized Coherent Structures and Periodic Solutions 
for the (2+1) dimensional K-dV and NNV equations}
\author{C. Senthil Kumar\thanksref{label1}},
\author{ R. Radha\thanksref{label2}},
\author{M. Lakshmanan
\thanksref{label1}\corauthref{cor1}}
\ead{lakshman@cnld.bdu.ac.in}
\corauth[cor1]{Corresponding Author.  Fax:+91-431-2407093}
%\thanks[label1]{}
%\thanks[label2]{}
\address[label1]{Centre for Nonlinear Dynamics, School of Physics, Bharathidasan University,
 Tiruchirapalli -620 024, India.}
\address[label2]{Department of Physics, Government College for Women,  \\
Kumbakonam - 612 001,
India.}
 
\begin{abstract}
In this paper, using a novel approach involving the truncated Laurent 
expansion in the Painlev\'e analysis of the (2+1) dimensional K-dV 
equation, we have trilinearized the evolution equation and obtained rather general classes of solutions in terms of arbitrary functions.  The highlight of this method is that it allows us to construct generalized periodic structures corresponding to different manifolds in terms of Jacobian elliptic functions, and the exponentially decaying dromions turn out to be special cases of these solutions.  We have also constructed multi-elliptic function solutions and multi-dromions and analysed their interactions.  The analysis is also extended to the case of generalized Nizhnik-Novikov-Veselov (NNV) equation, which is also trilinearized and general class of solutions obtained.
\end{abstract}
\end{frontmatter}

\section{Introduction}
The recent interest in the investigation of integrable models in (2+1) dimensions
can be attributed to the identification of dromions in the Davey-Stewartson I equation \cite{ref1,ref2}. 
These dromions which originate at the crosspoint of the intersection of two
nonparallel ghost solitons decay exponentially in all directions and are driven
by lower dimensional boundaries or velocity potentials (arbitrary functions) in the system.  It must
be mentioned that the presence of these so called boundaries enriches the
structure of (2+1) dimensional integrable models leading to the formation of
localized solutions like dromions, lumps, breathers, positons, etc.\cite {ref2,ref3}.  Thus, a
judicious harnessing of these lower dimensional arbitrary functions \cite {ref4,ref5,ref6,ref7,ref7_1,ref7_2} of space and
time may give rise to new spatio-temporal patterns in higher dimensions besides
throwing more light on their integrability.  

In this direction, we have recently identified a simple procedure to solve the (2+1) dimensional
systems, namely the Painlev\'e truncation approach (PTA) \cite{ref7}.
This approach converts the given evolution equation into a multilinear (in general) equation in terms of the noncharacteristic manifold.  This multilinear equation can be solved in terms of lower dimensional arbitrary functions of space and time.  Using this approach, one can generate various generalized periodic and localized solutions.  This was earlier demonstrated for the case of long wave-short wave resonance interaction equation in (2+1) dimensions by converting it into a trilinear equation \cite{ref7}.  In this paper, we wish to show that through the PTA the (2+1) dimensional K-dV equation equation can be trilinearized.  A four parameter special class of solutions involving three arbitrary functions $g(y)$, $h(y)$ and $f(x,t)$ in the indicated variables naturally arises for this trilinear equation.  This in turn leads to the universal form of solution obtained earlier by Tang, Lou and Zhang \cite{ref5} using a variable separation approach as a special case.  However, there exists a possibility of obtaining more general solution to the trilinear equation, which remains to be explored.  A large class of Jacobi elliptic function periodic solutions, multidromion and bound state solutions are also obtained.  It is also shown that the procedure can be extended to the case of Nizhnik-Novikov-Veselov (NNV) equation as well.

The plan of the paper is as follows.  In section 2, we show how the  
(2+1) dimensional K-dV equation can be trilinearized using the PTA and the solution obtained in terms of three arbitrary functions and four arbitrary parameters.
Sections  3 and 4 contain  two broad classes of solutions of the (2+1) dimensional K-dV equation, both periodic and exponentially localized ones, through judicious choices of the arbitrary functions. In section 5, we point out how the generalized NNV equation can be trilinearized through the PTA and how various periodic and localized solutions can be obtained.  Finally in section 6, we summarize our results.  The appendix contains the one dromion solution of the (2+1) dimensional K-dV equation obtained through Hirota bilinearization approach for comparison.

\section{The (2+1) dimensional K-dV equation and construction of solutions}
The (2+1) dimensional K-dV equation introduced by Boiti et al. \cite{ref8} has the form
\begin{equation}
q_{t}+q_{xxx}=3(q\partial_{y}^{-1}q_{x})_{x}. \label{1_1}
\end{equation}
This nonlocal equation (\ref{1_1}) reduces to the K-dV equation if $x=y$.    Introducing
a potential function $v(x,y,t)$ defined by the equation
\begin{equation}
v_{y}=q_{x},
\end{equation}
Eq.(\ref{1_1}) can be converted into a set of coupled equations of the form
\begin{subequations}
\begin{eqnarray}
q_{t}+q_{xxx}=3(qv)_{x},\label{1a} \\
v_{y}=q_{x}.\label{1b}
\end{eqnarray} \label{1}
\end{subequations}
The nature of (\ref{1b}) permits the presence of an arbitrary function $v'(x,t)$
as
\begin{equation}
v(x,y,t)=\int^{y}_{-\infty} q_x dy'+v'(x,t).
\end{equation} 
Expanding the physical field $q$ and the potential $v$ in the form of a Laurent
series in the neighbourhood of the noncharacteristic manifold $\phi(x,y,t)=0$,
($\phi_x=\phi_y \neq$ 0) and utilising the results of the Painlev\'e analysis \cite{ref9},
we obtain the following B\"acklund transformation by truncating the Laurent series at the constant
level term
\begin{subequations}
\begin{eqnarray}
q=q_0\phi^{-2}+q_1 \phi^{-1}+q_2, \\
v=v_0\phi^{-2}+v_1 \phi^{-1}+v_2, 
\end{eqnarray}\label{2}
\end{subequations}
where $(q,v)$ and $(q_2,v_2)$ are two different sets of solutions of the (2+1) dimensional K-dV
equation.  In ref. \cite{ref4} certain basic localized dromion solutions have been obtained using 
(\ref{2}) and Hirota bilinearization.  Here, we make use of the full power of Eq. (\ref{2})
by treating it as a variable transformation, and obtain rather general classes of solutions.
This new approach leads to a wide class of solutions not only for Eq. (\ref{1_1}) but for 
other (2+1) dimensional systems as well \cite{ref7}.

Considering now a vacuum solution of the form
\begin{equation}
q_2=0, v_2=v_2(x,t) \label{3}
\end{equation}
and substituting (\ref{3}) with (\ref{2}) in Eq. (\ref{1}), we obtain the following 
set of equations by equating the coefficients of $(\phi^{-5},\phi^{-3})$ to
zero,
\begin{subequations}
\begin{eqnarray}
-24 q_0 \phi_x^3 + 12 q_0 v_0 \phi_x = 0,\\
-v_0 \phi_y + q_0 \phi_x =0 .
\end{eqnarray}
\end{subequations}
Solving the above system of equations, we obtain 
\begin{subequations}
\begin{eqnarray}
q_0 = 2 \phi_x \phi_y, \\
v_0 = 2 \phi_x^2.
\end{eqnarray} \label{4}
\end{subequations}
Again, collecting the coefficients of $(\phi^{-4},\phi^{-2})$, we have
\begin{subequations}
\begin{eqnarray}
18 q_{0x} \phi_x^2+18 q_0 \phi_x \phi_{xx} -6q_1 \phi_x^3
-3(q_{0x} v_0-2 q_0 v_1 \phi_x & & \nonumber \\
-q_1 \phi_x v_0) 
-3(v_{0x} q_0-2 v_0 q_1 \phi_x - v_1 \phi_x q_0) &=& 0, \\
v_{0y}-v_1 \phi_y -(q_{0x}-q_1 \phi_x)&=&0.
\end{eqnarray}
\end{subequations}
Solving the above system of equations, we obtain
\begin{subequations}
\begin{eqnarray}
q_1 = -2 \phi_{xy}, \\
v_1 = -2 \phi_{xx}.
\end{eqnarray} \label{5}
\end{subequations}

Next, collecting the coefficients of $(\phi^{-3},\phi^{-1})$, we have
\begin{subequations}
\begin{eqnarray}
-2 q_0 \phi_t + (-6 q_{0xx} \phi_x-6q_{0x} \phi_{xx}-2 q_0 \phi_{xxx}
+6 q_{1x} \phi_x^2 & & \nonumber \\
+6 q_1 \phi_x \phi_{xx}) 
-3(q_{0x} v_1-2 q_0 \phi_x v_2 +q_{1x} v_0-q_1 v_1 \phi_x) & & \nonumber \\
-3(q_0 v_{1x}+q_1 v_{0x}-q_1v_1\phi_x)&=&0,  \label{6a} \\
v_{1y}-q_{1x} &=&0. \label{6b}
\end{eqnarray}
\end{subequations}
Here (\ref{6b}) is an identity as may be verified using (\ref{5}).
Solving (\ref{6a}) by using (\ref{4}) and (\ref{5}), we get the form of $v_2$ as
\begin{equation}
v_2= \frac{\phi_t+\phi_{xxx}}{3
\phi_x}+\frac{\phi_x\phi_{xxy}-\phi_{xx}\phi_{xy}}{\phi_x \phi_y}. \label{7}
\end{equation}

Now, we collect the coefficients of next order $(\phi^{-2},\phi^{0})$ to obtain
\begin{subequations}
\begin{eqnarray}
(q_{0t}-q_1 \phi_t)+(q_{0xxx}-3 q_{1x}\phi_{xx}-q_1 \phi_{xxx}-3
q_{1xx}\phi_x) & &\nonumber \\
-3(q_{0x}v_2+q_{1x}v_1-q_1 v_2 \phi_x)
-3(q_0 v_{2x} +q_1 v_{1x}) &=&0, \label{8a} \\
v_{2y}&=&0. \label{8b}
\end{eqnarray}
\end{subequations}
On the other hand, substituting the expression for the quantities $q_0$, $v_0$, $q_1$, $v_1$ and $v_2$ in Eq. (\ref{8a}), we obtain  the trilinear equation
\begin{equation}
\phi_{xx}(\phi_y \phi_{xyy}-\phi_{yy}\phi_{xy})+\phi_x (\phi_{yy}\phi_{xxy}-\phi_y\phi_{xxyy}) = 0. \label{8a_1}
\end{equation}
Substituting (\ref{7}) in (\ref{8b}), we obtain
\begin{eqnarray}
(\phi_{ty}+\phi_{xxxy})\phi_x\phi_y^2-(\phi_t+\phi_{xxx})\phi_{xy}\phi_y^2 
+3(\phi_x\phi_{xxyy}-\phi_{xx}\phi_{xyy})\phi_x \phi_y \nonumber \\
-3(\phi_x\phi_{xxy}-\phi_{xx}\phi_{xy})\phi_y\phi_{xy} 
-3(\phi_x\phi_{xxy}-\phi_{xx}\phi_{xy})\phi_x\phi_{yy} =0. \label{8b_2} 
\end{eqnarray}
Using Eq. (\ref{8a_1}) into (\ref{8b_2}), the later reduces to the form
\begin{eqnarray}
(\phi_{ty}+\phi_{xxxy})\phi_x\phi_y-(\phi_t+\phi_{xxx})\phi_{xy}\phi_y 
-3(\phi_x\phi_{xxy}-\phi_{xx}\phi_{xy})\phi_{xy} =0. \label{8b_1} 
\end{eqnarray}
which is again in a trilinear form.  Thus equations (\ref{8a_1}) and (\ref{8b_1}) may be considered as  the equivalent trilinear forms of Eqs. (\ref{1}).

One can immediately observe that the set of trilinear equations (\ref{8a_1}) and (\ref{8b_1}) admit a set of two arbitrary functions and their products as solutions
\begin{subequations}
\begin{eqnarray}
 & \phi = f(x,t), \\
 & \phi = g(y) \\
\mbox{and} &  \nonumber \\
 & \phi = f(x,t)g(y),
\end{eqnarray}
\end{subequations}
where $f$ and $g$ are arbitrary functions in the indicated 
variables.  In fact a more general solution involving three arbitrary functions and four arbitrary constant parameters can be easily identified:
\begin{equation}
\phi(x,y,t)=c_1+c_2 f(x,t)+c_3 h(y)+c_4 f(x,t)g(y), \label{8c}
\end{equation}
where $h(y)$ and $g(y)$ are in general different arbitrary functions of $y$. Here $c_1$, $c_2$, $c_3$, and $c_4$ are arbitrary parameters.  The problem of identifying even more general solution to (\ref{8a_1}) and (\ref{8b_1}) remains to be investigated, while we will concentrate on the special solution (\ref{8c}) in this paper.

Collecting the coefficients $(\phi^{-1}, \phi)$, we obtain
\begin{equation}
q_{1t}+q_{1xxx}-3q_{1x}v_2-3q_1v_{2x}=0. \label{11}
\end{equation}
It can be checked that equation (\ref{11}) is compatible with the earlier results. 

Now substituting the above form (\ref{8c}) for the manifold $\phi(x,y,t)$ into the truncated Painlev\'e series (\ref{2}) for the functions $q(x,y,t)$ and $v(x,y,t)$, along with the expressions for the coefficient functions $q_0$, $v_0$, $q_1$, $v_1$ and $v_2$ given above, we finally obtain the solution to Eqs. (\ref{1}) as
%\begin{subequations}
\begin{eqnarray}
 q(x,y,t)&=&\frac{2 f_x[(c_2+c_4g)c_3h_y-(c_1+c_3h)c_4g_y]}{[c_1+c_2 f(x,t)+c_3 h(y)+c_4 f(x,t)g(y)]^2} \label {8c_1}, \\
v(x,y,t) &=&\frac{2 (c_2+c_4g)^2 f_x^2}{[c_1+c_2 f(x,t)+c_3 h(y)+c_4 f(x,t)g(y)]^2} \nonumber \\
& & -\frac{2 (c_2+c_4g) f_{xx}}{c_1+c_2 f(x,t)+c_3 h(y)+c_4 f(x,t)g(y)}+\frac{f_t+f_{xxx}}{3f_x}.  \label{8c_2}
\end{eqnarray}
%\end{subequations}
It may be noted that the expressions given in (\ref{8c_1}) and (\ref{8c_2}) coincide exactly with the forms obtained by Tang, Lou and Zhang \cite{ref5} using the method of variable separation for the special case $g(y)=h(y)$ in Eq. (\ref{8c}).  Another special case $c_1=c_2=0$, $c_3=c_4=1$ was identified by Peng recently \cite{ref7_2}.  Also, in our case it is clear that finding any solution to (\ref{8a_1}) and (\ref{8b_1}) which is more general than (\ref{8c}) will lead to more general solution than (\ref{8c_1})-(\ref{8c_2}).  This problem is being investigated further.  In the following, we will investigate the nature of solutions (\ref{8c_1})-(\ref{8c_2}) corresponding to periodic and localized solutions.  For this purpose, we shall consider the special cases (i) $c_4=0$ and (ii) $c_2=c_3=0$ in (\ref{8c}) or (\ref{8c_1})-(\ref{8c_2}), corresponding to the sum and product of arbitrary functions, respectively, and investigate the nature of periodic and localized solutions.  It may be noted that there is no difficulty in proceeding with the general form (\ref{8c_1})-(\ref{8c_2}) also except for the fact that the expression will be more lengthy which we desist from presenting here.

\section{Periodic and localized solutions corresponding to sum of arbitrary functions}
In this section, we will assume $c_1=c$, $c_2=c_3=1$, $c_4=0$ in (\ref{8c}) or (\ref{8c_1})-(\ref{8c_2}). Consequently, we have   
\begin{subequations}
\begin{eqnarray}
q&=&\frac{2 f_x h_y}{(f+h+c)^2}, \label{S1} \\
v&=&\frac{2 f_x^2}{(f+h+c)^2}-\frac{2 f_{xx}}{(f+h+c)}
+\frac{f_t+f_{xxx}}{3f_x}.  \label{S2}
\end{eqnarray}\label{S3}
\end{subequations}

\subsection{Harnessing of arbitrary functions and novel solutions of (2+1) dimensional K-dV equation}
Let us now choose the arbitrary functions $f(x,t)$ and $h(y)$ to be  Jacobian elliptic
functions, namely $sn$ or $cn$ functions.  The motivation behind this choice of
arbitrary function stems from the fact that the limiting forms of these
functions happen to be localized functions.  Hence, a choice
of $cn$ and $sn$ functions can yield periodic solutions which are more generalized
than exponentially localized solutions (dromions).  We choose, for example,
\begin{equation}
f=\alpha\, \mbox{sn}(a x+c_1 t+d_1;m_1),\,h=\beta\, \mbox{sn}(by+d_2;m_2) \label{12}
\end{equation}
so that
\begin{equation}
q(x,y,t)=\frac{2\alpha\,\beta \,a\,b \,\mbox{cn}(u_1;m_1) \mbox{dn}(u_1;m_1) 
 \,\mbox{cn}(u_2;m_2) \mbox{dn}(u_2;m_2)}
{(c+\alpha \mbox{sn}(u_1;m_1)+\beta \mbox{sn}(u_2;m_2))^2}  \label{13},
\end{equation}
where $u_1=ax + c_1 t + d_1$ and $u_2=by+d_2$.
In Eqs. (\ref{12}) and (\ref{13}), the quantities $m_1$ and $m_2$ are moduli of the respective Jacobian elliptic functions while $\alpha$, $\beta$, 
$a$, $b$, $c$, $c_1$, $d_1$ and $d_2$ are arbitrary constants.  We choose the constant parameters $c$, $\alpha$ and $\beta$ such that $|c|>|\alpha+\beta|$ for nonsingular solution.
The profile of the above solution for the parametric choice $\alpha=\beta$=1, $a$=0.5, $b$=0.4, $c$=-4,
$c_1$=-1, 
$d_1$=$d_2$=0,  $m_1$= 0.2, $m_2$ = 0.3 at time $t$=0 is shown in Fig.1(a).
Note that the periodic wave travels along the x-direction only. \\

\subsubsection{(1,1) dromion solution}
As a limiting case of the periodic solution given by  Eq. (\ref{13}), 
when $m_1$, $m_2$ $\rightarrow$ 1, the above 
solution degenerates into an exponentially localized solution (dromion).
Noting that $cn(u;1)=dn(u;1)=\mbox{sech}u$ and $sn(u;1)=\mbox{tanh}u$, 
the limiting form corresponding to (1,1) dromion takes the expression
\begin{equation}
q=\frac{2a\,b\,\alpha\,\beta\, \mbox{sech}^2(ax + c_1 t + d_1) \mbox{sech}^2(by+d_2)}
{(c+\alpha\mbox{tanh}(ax + c_1 t + d_1)+\beta \mbox{tanh}(by+d_2))^2}, \;\; (|c|>|\alpha+\beta|) \label{14}
\end{equation}
and the variable $v$ then takes the form (using expression (\ref{S2}))
\begin{eqnarray}
v&=&\frac{2\alpha^2\,a^2\, \mbox{sech}^4(ax + c_1 t + d_1)}
{(c+\alpha\, \mbox{tanh}(ax + c_1 t + d_1)+\beta \mbox{tanh}(by+d_2))^2} \nonumber \\
&&+\frac{4\alpha\,a^2\,\mbox{sech}^2(ax + c_1 t + d_1)\mbox{tanh}(ax + c_1 t + d_1)} 
{(c+\alpha\,\mbox{tanh}(ax + c_1 t + d_1)+\beta \mbox{tanh}(by+d_2))}+ \nonumber \\
&&\frac{1}{3 a}[c_1+4 a^3 \mbox{tanh}^2(ax + c_1 t + d_1)-2 a^3 \mbox{sech}^2(ax + c_1 t + d_1)]. \label{15}
\end{eqnarray}
Schematic form of the (1,1) dromion for the parametric choice $\alpha=\beta$=1, $a$=0.5, $b$=0.4, $c$=-4,
$c_1$=-1, $d_1$=$d_2$=0 at time $t$=0  is shown in Fig.1(b). Again note that
the dromion travels along the x-direction. One can obtain the (1,1) dromion given by Radha and Lakshmanan in ref. \cite{ref4} by fixing the parameters $\alpha$, $\beta$, $a$, $b$, $c_1$, $d_1$, $d_2$  of equation (\ref{14}) suitably (see also the Appendix).

\subsubsection{More general periodic solution and (2,1) dromion}
Next we obtain a more general periodic solution by choosing further general 
forms for the arbitrary functions.  As an example, we choose
%\begin{subequations}
\begin{eqnarray}
f&=& \alpha_1 \mbox{sn}(a_1 x+c_1 t+d_1;m_1)+\alpha_2 \mbox{sn}(a_2 x+c_2 t+d_2;m_2),\nonumber \\
h&=&\beta \mbox{sn}(by+d_3;m_3), \label{16}
\end{eqnarray} 
%\end{subequations}
where $b$, $\beta$, $\alpha_i$, $a_i$,  $c_i$ and $d_j$ are arbitrary constants ($i=1,2$; $j=1,2,3$) and $m_j$'s are modulus parameters.
Then
\begin{equation}
q= \frac{q_1}{q_2},  \label{17}
\end{equation}
where, $q_1=2[(\alpha_1 a_1\mbox{cn}(u_1;m_1)\mbox{dn}(u_1;m_1)+\alpha_2 a_2\mbox{cn}(u_2;m_2)\mbox{dn}(u_2;m_2))$\\ $
\beta\,b \, \mbox{cn}(u_3;m_3)\mbox{dn}(u_3;m_3)]$, $q_2=(c+\alpha_1 \mbox{sn}(u_1;m_1)+\alpha_2 \mbox{sn}(u_2;m_2)+\beta\mbox{sn}(u_3;m_3))^2$, $u_1=a_1 x+c_1 t+d_1$, $u_2=a_2 x+c_2 t+d_2$ and $u_3=by+d_3$ with corresponding expressions for $v(x,y,t)$.  We choose the constant $c$ such that $|c|>|\alpha_1+\alpha_2+\beta|$ for nonsingular solutions.
The profile of the above solution for the parametric choice $\alpha_1=\alpha_2=\beta$=1, $a_1$ =0.5,
$a_2$ =0.5, $b$ =0.4, $c$=-4, $c_1$=-1, $c_2$=-2, $d_1$=$d_2$=$d_3$=0,  $m_1$= 0.2, $m_2$ = 0.3, $m_3$ = 0.4 
at time $t$=0 is shown in Fig.2a.  As  $m_1$, $m_2$, $m_3$ $\rightarrow$ 1, the above solution given by Eq. (\ref{17}) degenerates into a (2,1) dromion solution given by
\begin{equation}
q= \frac{2(\alpha_1 a_1\mbox{sech}^2u_1+\alpha_2 a_2\mbox{sech}^2u_2)
b\,\mbox{sech}^2u_3}
{(c+\alpha_1 \mbox{tanh}u_1+\alpha_2 \mbox{tanh}u_2+\beta \mbox{tanh}u_3)^2}, \;\; (|c|>|\alpha_1+\alpha_2+\beta|)\label{18}
\end{equation}
where $u_1=a_1 x+c_1 t+d_1$, $u_2=a_2 x+c_2 t+d_2$ and $u_3=by+d_3$.  The solution is plotted for the parametric choice 
$\alpha_1=\alpha_2=\beta$=1, $a_1$ =0.5, $a_2$= 0.8, $b$ =0.4, $c$=-4, $c_1$=-1, $c_2$=-2.5, $d_1$=$d_2$=$d_3$=0, for various values of $t$ in Figs. 2(b)-2(d) in order to bring out the dromion interaction clearly.

\subsubsection{Asymptotic analysis for (2,1) dromion solution}
Since both the dromions in Figs. 2(b)-2(d) corresponding to the solution (\ref{18}) are travelling along the x-direction, following one another, it is enough to do this analysis for $y=0$ (and  $d_3=0$ so that $u_3=0$). 
Similar analysis holds good for any other value of $y$ and $d_3 \neq 0$. This restriction corresponds to the cross section of dromions, which are essentially solitons.  We analyse the limits $t$ $\rightarrow$ $-\infty$ and $t$ $\rightarrow$ $+\infty$ separately so as to understand the interaction of dromions centered around $u_1 \approx 0$ or $u_2 \approx 0$.  Without loss of generality, let us assume $c_1>c_2$ and $a_1<a_2$.  Then, we find in the limit $t$ $\rightarrow$ $\pm\infty$, $u_1=a_1 x+c_1 t+d_1$ and $u_2=a_2 x+c_2 t+d_2$ take the following limiting values.

1) As t $\rightarrow$ $-\infty$:
\begin{eqnarray} 
    u_1 \approx 0, u_2 \rightarrow +\infty \nonumber \\
    u_2 \approx 0, u_1 \rightarrow -\infty \nonumber 
\end{eqnarray}

2) As t $\rightarrow$ $+\infty$:
\begin{eqnarray} 
    u_1 \approx 0, u_2 \rightarrow -\infty \nonumber \\
    u_2 \approx 0, u_1 \rightarrow +\infty \nonumber 
\end{eqnarray}

1. Before interaction (as t $\rightarrow$ $-\infty$): \\
For $u_1 \approx 0, u_2 \rightarrow +\infty$ and $\alpha_1=\alpha_2=\beta=1$, the (2,1) dromion solution (\ref{18}) becomes (soliton solution corresponding to dromion 1) 
\begin{equation}
q = \frac{2 a_1 b}{c(c+2)}\mbox{sech}^2(u_1+\delta_1), \;\;\;\delta_1=\sqrt{\frac{c+2}{c}}. 
\end{equation}
For $u_2 \approx 0, u_1 \rightarrow -\infty$, (\ref{18}) becomes 
(soliton solution corresponding to dromion 2) 
\begin{equation}
q = \frac{2 a_2 b}{c(c-2)}\mbox{sech}^2(u_2+\delta_2), \;\;\;\delta_2=\sqrt{\frac{c}{c-2}}. 
\end{equation}

2. After interaction (as t $\rightarrow$ $+\infty$): \\
For $u_1 \approx 0, u_2 \rightarrow -\infty$,  (\ref{18}) becomes 
(soliton solution corresponding to dromion 1) 
\begin{equation}
q = \frac{2 a_1 b}{c(c-2)}\mbox{sech}^2(u_1+\delta_2). 
\end{equation}
For $u_2 \approx 0, u_1 \rightarrow +\infty$,  (\ref{18}) becomes 
(soliton solution corresponding to dromion 2) 
\begin{equation}
q = \frac{2 a_2 b}{c(c+2)}\mbox{sech}^2(u_2+\delta_1). 
\end{equation}

The above results can be interpreted  in the following way.  Before interaction, the dromion 1 has a larger amplitude, travelling slower and the dromion 2 has a shorter amplitude, travelling faster.  During interaction (at time $t=0$), an exchange of energy between the dromions take place.  This results in the gain in amplitude of dromion 2 and fall in amplitude of dromion 1.  But there is no change in velocity of the dromions and there is only a change in phase.

\subsubsection{Bounded multiple solitary waves}
One can also construct bounded two solitary waves by choosing
%\begin{subequations}
\begin{eqnarray}
f&=&\alpha\, \mbox{sn}(a x+c_1 t+d_1;m_1),\nonumber \\
h&=&\beta_1\, \mbox{sn}(b_1 y+d_2;m_2)+\beta_2\, \mbox{sn}(b_2 y+d_3;m_3), \label{19}
\end{eqnarray} 
%\end{subequations}
where $\alpha$, $a$, $c_1$, $\beta_i$, $b_i$, and $d_j$ ($i=1,2$; $j=1,2,3$) are arbitrary constants and $m_j$'s are modulus parameters.
Then
\begin{equation}
q= \frac{q_3}{q_4},  \label{20}
\end{equation}
where $q_3=2\alpha\, a\,\mbox{cn}(u_1;m_1)\mbox{dn}(u_1;m_1)
[\beta_1 b_1 \mbox{cn}(u_2;m_2)\mbox{dn}(u_2;m_2)+\beta_2 b_2 \mbox{cn}(u_3;m_3)$ $\mbox{dn}(u_3;m_3)]$,
$q_4=(c+\alpha \mbox{sn}(u1;m1)+\beta_1 \mbox{sn}(u2;m2)+\beta_2 \mbox{sn}(u3;m3))^2$, $u_1=a_1 x+c_1 t+d_1$, $u_2=b_1y+d_2$ and $u_3=b_2 y+d_3$ with corresponding expressions for $v(x,y,t)$.  Here $|c|>|\alpha+\beta_1+\beta_2|$.
The profile of the above solution for the parametric choice $\alpha=\beta_1=\beta_2=1$, $a$ =0.5,
$b_1$ =2.5, $b_2$ =1.7, $c$=-4, $c_1$=1, $c_2$=-2, $d_1$=$d_2$=$d_3$=0,  $m_1$= 0.2, $m_2$ = 0.3, $m_3$ = 0.4, $t$=0 is shown in
Fig.3(a).  
As  $m_1$, $m_2$, $m_3$ $\rightarrow$ 1, the above solution, Eq. (\ref{20}),
degenerates into a
bounded two dromion solution given by
\begin{equation}
q= \frac{2\alpha\,a\,\mbox{sech}^2u_1
(\beta_1 b_1 \mbox{sech}^2u_2+\beta_2 b_2 \mbox{sech}^2u_3)}
{(c+\alpha \mbox{tanh}u_1+\beta_1 \mbox{tanh}u_2+\beta_1 \mbox{tanh}u_3)^2}, \label{21}
\end{equation}
where $u_1=a x+c_1 t+d_1$, $u_2=b_1y+d_2$ and $u_3=b_2 y+d_3$.
 The above solution for the parametric choice 
$\alpha=\beta_1=\beta_2=1$, $a$ =0.5, $b_1$ =2.5, $b_2$ =1.7, $c$=-4, $c_1$=1, $c_2$=-2, $d_1$=$d_2$=$d_3$=0,
is shown in Fig.3(b).  Here both the dromions travel with equal velocity along the
$x$ direction. Since they move parallel to each other there is no interaction
between them.

\subsection{(N,M) dromion solution}
Proceeding in a similar way as above, one can generate (N,M) dromion solution by choosing
\begin{subequations}
\begin{eqnarray}
f=\sum_{i=1}^N \alpha_i \mbox{sn}(a_i x+c_i t+d_{1i};m_i), \\
h=\sum_{j=1}^M \beta_i \mbox{sn}(b_j y+d_{2j};m_j), 
\end{eqnarray} 
\end{subequations}
where $\alpha_i$, $\beta_i$, $a_i$, $b_j$, $c_i$, $d_{1i}$, $d_{2j}$ are arbitrary constants and $m_i$, $m_j$ are modulus parameters and substituting them in the expression for $q$ and $v$ in (\ref{S3}).  As the analysis is similar to the above, we do not present further details here.

\subsection{Singular solutions}

While presenting the explicit dromion solutions above, we have always assumed the magnitude of the parameters $c$ in the solution (\ref{S3}) to be greater than a constant value in order to have nonsingular solutions.  For example, for the (1,1) dromion solution (\ref{14}), we have taken $|c|>|\alpha+\beta|$.  For $|c| \le |\alpha+\beta|$, the solution (\ref{14}) becomes singular and it is illustrated in Figs.4(a)-4(c) for different instants of time.  Here we have chosen  $\alpha=\beta=1$, $c=1$, $a$ = 0.5, $b$ = 0.4, $c$ = 1, $d_1$ = $d_2$ = 0, $\alpha_1=\alpha_2=1$.

\section{Solutions corresponding to product of arbitrary functions}
In this section, we will assume $c_2=c_3=0$ and $c_4=1$ in (\ref{8c}) or (\ref{8c_1})-(\ref{8c_2})
and call $c_1=c$. Then the solution can be written as   
\begin{subequations}
\begin{eqnarray}
q&=&\frac{-2c f_x g_y}{(f\,g+c)^2}, \label{S4} \\
v&=&\frac{2 f_x^2 g^2}{(f\,g+c)^2}-\frac{2 f_{xx}g}{(f\,g+c)}
+\frac{f_t+f_{xxx}}{3f_x}.  \label{S5}
\end{eqnarray}\label{S6}
\end{subequations}
Now we will look for periodic and localized solutions.

\subsection{Harnessing of arbitrary functions and novel solutions of (2+1) dimensional K-dV equation}
As before, we choose the arbitrary functions $f$ and $g$ to be  Jacobian elliptic
functions, namely $sn$ or $cn$ functions,  
\begin{equation}
f=\alpha \mbox{sn}(a x+c_1 t+d_1;m_1),\,g=\beta \mbox{sn}(by+d_2;m_2) \label{23}
\end{equation}
so that
\begin{equation}
q(x,y,t)=\frac{q_5}{q_6}  \label{24},
\end{equation}
where $q_5=-2\,c \,\alpha\,\beta\, a \,b \,\mbox{cn}(u_1;m_1) \mbox{dn}(u_1;m_1) 
 \,\mbox{cn}(u_2;m_2) \mbox{dn}(u_2;m_2)$, $q_6=(c+\alpha\,\beta\,\mbox{sn}$ $(u_1;m_1)\mbox{sn}(u_2;m_2))^2$, $u_1=ax + c_1 t + d_1$ and $u_2=by+d_2$.
In Eqs. (\ref{23}) and (\ref{24}), the quantities $m_1$ and $m_2$ are the
moduli of the respective Jacobian elliptic functions while
$\alpha$, $\beta$, $a$, $b$, $c$, $c_1$, $d_1$ and $d_2$ are arbitrary constants with $|c|>|\alpha\beta|$.
The profile of the above solution for the parametric choice $\alpha=\beta=1$, $a$=0.5, $b$=0.4, $c$=-4,
$c_1$=-1, 
$d_1$=$d_2$=0,  $m_1$= 0.2, $m_2$ = 0.3 at time $t$=0 is shown in Fig. 5(a).
It can be observed that here again the periodic wave moves along the $x$-direction only.\\

\subsubsection{(1,1) dromion solution}
As a limiting case of the periodic solution given by  Eq. (\ref{24})
when $m_1$, $m_2$ $\rightarrow$ 1, the above 
solution degenerates into an exponentially localized solution (dromion).
The limiting form corresponding to the (1,1) dromion take the expression
\begin{equation}
q=\frac{-2 c\,\alpha\, \beta\, a\,b\, \mbox{sech}^2(ax + c_1 t + d_1) \mbox{sech}^2(by+d_2)}{(c+\alpha\, \beta\,\mbox{tanh}(ax + c_1 t + d_1)\mbox{tanh}(by+d_2))^2},\label{25}
\end{equation}
with $|c|>|\alpha\beta|$ and the variable $v$ then takes the form (using expression (\ref{S5}))
\begin{eqnarray}
v&=&\frac{2a^2\alpha^2\, \beta^2\, \mbox{sech}^4(ax + c_1 t + d_1)\mbox{tanh}^2(by+d_2)}
{(c+\alpha\, \beta\,\mbox{tanh}(ax + c_1 t + d_1)\mbox{tanh}(by+d_2))^2} \nonumber \\
&&+\frac{4a^2\alpha\, \beta\,\mbox{sech}^2(ax + c_1 t + d_1)\mbox{tanh}(ax + c_1 t + d_1)\mbox{tanh}(by+d_2)} 
{(c+\alpha\, \beta\,\mbox{tanh}(ax + c_1 t + d_1)\mbox{tanh}(by+d_2))}+ \nonumber \\
&&\frac{1}{3 a}[c_1+4 a^3 \mbox{tanh}^2(ax + c_1 t + d_1)
-2 a^3 \mbox{sech}^2(ax + c_1 t + d_1)]. \label{26}
\end{eqnarray}
Schematic form of the (1,1) dromion for the parametric choice $\alpha=\beta=1$, $a$=0.5, $b$=0.4, $c$=-4,
$c_1$=-1, $d_1$=$d_2$=0 at time $t$=0  is shown in Fig.5(b). Again note that
the dromion travels along the x-direction.

\subsubsection{More general periodic solution and (2,1) dromion}
Next, we obtain more general periodic solution by choosing further general 
forms for the arbitrary functions.  As an example, we choose
%\begin{subequations}
\begin{eqnarray}
f&=& \alpha_1 \mbox{sn}(a_1 x+c_1 t+d_1;m_1)+\alpha_2 \mbox{sn}(a_2 x+c_2 t+d_2;m_2),\nonumber \\
g&=& \beta\mbox{sn}(by+d_3;m_3), \label{27}
\end{eqnarray} 
%\end{subequations}
where $b$, $\beta$, $\alpha_i$, $a_i$, $c_i$ and $d_j$ are arbitrary constants ($i=1,2$; $j=1,2,3$) and $m_j$'s are modulus parameters.
Then
\begin{equation}
q= \frac{q_9}{q_{10}},  \label{28}
\end{equation}
where $q_9=-2c\,\beta\,[(\alpha_1 a_1\mbox{cn}(u_1;m_1)\mbox{dn}(u_1;m_1)+\alpha_2 a_2\mbox{cn}(u_2;m_2)\mbox{dn}(u_2;m_2))$\\ $
b \,\mbox{cn}(u_3;m_3)\mbox{dn}(u_3;m_3)]$, $q_{10}=[c+\beta (\alpha_1 \mbox{sn}(u_1;m_1)+\alpha_2 \mbox{sn}(u_2;m_2))
\mbox{sn}(u_3;m_3)]^2$, $u_1=a_1 x+c_1 t+d_1$, $u_2=a_2 x+c_2 t+d_2$ and $u_3=by+d_3$ with corresponding expressions for $v(x,y,t)$.  Here $|c|>|(\alpha_1+\alpha_2)\beta|$.
The profile of the above solution for the parametric choice $\alpha_1=\alpha_2=\beta=1$, $a_1$ =0.5,
$a_2$ =0.8, $b$ =0.4, $c$=-4, $c_1$=-1, $c_2$=-2.5, $d_1$=$d_2$=$d_3$=0,   $m_1$= 0.2, $m_2$ = 0.3, $m_3$ = 0.4 
at time $t$=0 is shown in Fig.6(a).  As  $m_1$, $m_2$, $m_3$ $\rightarrow$ 1, the above solution Eq. (\ref{28})
degenerates into a (2,1) dromion solution given by
\begin{equation}
q= \frac{-2c\,b \,\beta (\alpha_1 a_1\mbox{sech}^2u_1+\alpha_2 a_2\mbox{sech}^2u_2)
\,\mbox{sech}^2u_3}
{(c+\beta (\alpha_1 \mbox{tanh}u_1+\alpha_2 \mbox{tanh}u_2)\mbox{tanh}u_3)^2}, \label{29}
\end{equation}
$|c|>|(\alpha_1+\alpha_2)\beta|$, where $u_1=a_1 x+c_1 t+d_1$, $u_2=a_2 x+c_2 t+d_2$ and $u_3=by+d_3$.
 The dromion interaction for the parametric choice $\alpha_1=\alpha_2=\beta=1$,
$a_1$ =$a_2$ =0.5,$b$ =0.4, $c$=-4, $c_1$=-1, $c_2$=-2, $d_1$=$d_2$=$d_3$=0,
is shown in Figs.6(b)-6(d) for different time
intervals.  
\subsubsection{Asymptotic analysis for (2,1) dromion solution corresponding to Eq.(\ref{29})}
Proceeding with the analysis as was done in the previous section,  we find that both the dromions are of
different amplitudes and travelling with different velocities along the x-direction.
At time $t=0$, there is an interaction between the dromions. After interaction, we find that there is no change in amplitude or velocity of the dromions .

\subsubsection{Bounded multiple solitary waves}
One can also construct bounded two solitary waves as before by choosing
%\begin{subequations}
\begin{eqnarray}
f&=&\alpha \,\mbox{sn}(a x+c_1 t+d_1;m_1),\nonumber \\
g&=&\beta_1 \mbox{sn}(b_1 y+d_2;m_2)+\beta_2 \mbox{sn}(b_2 y+d_3;m_3), \label{30}
\end{eqnarray} 
%\end{subequations}
where $a$, $\alpha$,  $c_1$, $b_i$ and $d_j$ are arbitrary constants ($i=1,2$; $j=1,2,3$) and $m_j$'s are modulus parameters.
Then
\begin{equation}
q= \frac{q_{11}}{q_{12}},  \label{31}
\end{equation}
where $q_{11}=-2c\, a\,\alpha\,\mbox{cn}(u_1;m_1)\mbox{dn}(u_1;m_1)
[\beta_1 b_1 \mbox{cn}(u_2;m_2)\mbox{dn}(u_2;m_2)+\beta_2 b_2 \mbox{cn}(u_3;$ $m_3)\mbox{dn}(u_3;m_3)]$,
$q_{12}=(c+\alpha\,\mbox{sn}(u_1;m_1)[\beta_1 \mbox{sn}(u_2;m_2)+\beta_2 \mbox{sn}(u_3;m_3)])^2$, $u_1=a_1 x+c_1 t+d_1$, $u_2=b_1y+d_2$ and $u_3=b_2 y+d_3$ with corresponding expressions for $v(x,y,t)$.
Again here $|c|>|\alpha(\beta_1+\beta_2)|$ for nonsingular solutions.
The profile of the above solution for the parametric choice $\alpha=\beta_1=\beta_2=1$, $a$ =0.5,
$b_1$ =2.5, $b_2$ =1.7, $c$=-4, $c_1$=1, $c_2$=-2, $d_1$=$d_2$=$d_3$=0,  $m_1$= 0.2, $m_2$ = 0.3, $m_3$ = 0.4, $t$=0 is shown in Fig.8a.  As  $m_1$, $m_2$, $m_3$ $\rightarrow$ 1, the above solution given by Eq. (\ref{31})
degenerates into a
bounded two dromion solution given by
\begin{equation}
q= \frac{-2c\,\alpha\, a\,\mbox{sech}^2u_1
(\beta_1 b_1 \mbox{sech}^2u_2+\beta_2 b_2 \mbox{sech}^2u_3)}
{(c+\alpha\,\mbox{tanh}u_1(\beta_1 \mbox{tanh}u_2+\beta_2 \mbox{tanh}u_3))^2}, \label{32}
\end{equation}
$|c|>|\alpha(\beta_1+\beta_2)|$, where $u_1=a x+c_1 t+d_1$, $u_2=b_1y+d_2$ and $u_3=b_2 y+d_3$ with the parametric choice being
$\alpha=\beta_1=\beta_2=1$, $a$ =0.5, $b_1$ =2.5, $b_2$ =1.7, $c$=-4, $c_1$=1, $c_2$=-2, $d_1$=$d_2$=$d_3$=0.  The (2,1) dromion 
is shown in Fig.7(b).  The two dromions evolve quite similar to the pattern of Fig.3(b) in the x-direction.

\section{Generalized Nizhnik-Novikov-Veselov (NNV) Equation and construction of solutions}
The generalized Nizhnik-Novikov-Veselov (NNV) equation is a symmetric generalization of the K-dV equation in (2+1) dimensions and is given by
\begin{subequations}
\begin{eqnarray}
u_t+a u_{xxx}+b u_{yyy}+c u_x+d u_y-3av_x u-3 a v u_x& & \nonumber \\
-3 b w_y u-3b w u_y&=&0,\label{} \\
u_x&=&v_y,\label{} \\
u_y&=&w_x.
\end{eqnarray} \label{39}
\end{subequations}
Here $a$, $b$, $c$ and $d$ are parameters.
This equation which is also known to be completely integrable has been investigated and exponentially localized solutions have been generated \cite{ref4,ref7_1}.  We now apply the  Painlev\'e truncation approach to obtain more general solutions.  For this purpose, we again truncate the Laurent series at the constant level term to get the transformation
\begin{subequations}
\begin{eqnarray}
u=u_0\phi^{-2}+u_1 \phi^{-1}+u_2, \\
v=v_0\phi^{-2}+v_1 \phi^{-1}+v_2, \\
w=w_0\phi^{-2}+w_1 \phi^{-1}+w_2.
\end{eqnarray}\label{38}
\end{subequations}
Again considering the vacuum solutions of the form
\begin{equation}
u_2=0, v_2 = v_2(x,t), w_2 = w_2(y,t),
\end{equation}
where $v_2(x,t)$ and $w_2(y,t)$ are arbitrary functions in the indicated variables,
and collecting the coefficients of different powers of $\phi$ as before and solving the resultant equations, we obtain the following results:
\begin{subequations}
\begin{eqnarray}
u_0=2\phi_x\phi_y,\;\;\; u_1=-2\phi_{xy}, \\
v_0=2\phi_x^2,\;\;\; v_1=-2\phi_{xx}, \\
w_0=2\phi_y^2,\;\;\; w_1=-2\phi_{yy}. 
\end{eqnarray}\label{37}
\end{subequations}
Also $v_2$ and $w_2$ are related through $\phi$ by the relation
\begin{eqnarray}
v_2 &=& \frac{\phi_t+a \phi_{xxx}+ b \phi_{yyy}}{3a
\phi_x}+\frac{(\phi_x\phi_{xxy}-\phi_{xx}\phi_{xy})}{\phi_x \phi_y}
+\frac{b(\phi_y\phi_{xyy}-\phi_{yy}\phi_{xy})}{a\phi_x^2}  \nonumber \\
& & +\frac{c}{3a}+\frac{d\phi_y}{3a \phi_x}
-\frac{b \phi_y w_2}{a \phi_x}. \label{41}
\end{eqnarray}
Making use of the expression (\ref{37}) and (\ref{41}) in (\ref{38}) and using the resultant forms in the generalized NNV equation (\ref{39}), we obtain the following system of equations which is triliner in $\phi$ (with the coefficient $w_2$ also present),
\begin{subequations}
\begin{eqnarray}
\phi_{xx}(\phi_y \phi_{xyy}-\phi_{yy}\phi_{xy})+\phi_x (\phi_{yy}\phi_{xxy}-\phi_y\phi_{xxyy}) &=& 0, \label{40_1} \\
\phi_x\phi_y(\phi_{yt}+a \phi_{yxxx}+b \phi_{yyyy})-\phi_y(\phi_t+a \phi_{xxx}+b \phi_{yyy})\phi_{xy} & &\nonumber \\
-3a(\phi_x\phi_{xxy}-\phi_{xx}\phi_{xy})\phi_{xy}+3b\phi_y(\phi_y\phi_{xyyy}-\phi_{yyy}\phi_{xy}) & &\nonumber \\
+6b\phi_y(\phi_{yy}\phi_{xyy}-\phi_y\phi_{xxyy})+d\phi_y(\phi_x\phi_{yy}-\phi_y\phi_{xy}) & &\nonumber \\
-3b\phi_y[\phi_x(\phi_{yy}w_2+\phi_yw_{2y})-\phi_y\phi_{xy}w_2]&=&0, \label{40_2} \\
w_{2x}&=&0. \label{40_3}
\end{eqnarray}\label{40}
\end{subequations}
One can observe that the above set of equations (\ref{40}) admits a more general form of $\phi$ involving two
arbitrary functions $f(x,t)$ and $g(y,t)$, and four arbitrary parameters, $c_1$, $c_2$, $c_3$ and $c_4$ as
\begin{equation}
\phi(x,y,t)=c_1+c_2 f(x,t)+c_3 g(y,t)+c_4 f(x,t)g(y,t). \label{42}
\end{equation}
Here $v_2$ and $w_2$ take the forms (from(\ref{41}))
\begin{subequations}
\begin{eqnarray}
v_2 &=& \frac{f_t+a f_{xxx}+c f_x+(c_3+c_4 f)}{3a f_x}, \\
w_2 &=& \frac{g_t+b g_{yyy}+d g_y-(c_2+c_4 g)}{3b g_y}.
\end{eqnarray}\label{42_1}
\end{subequations}
Here $c_1$, $c_2$, $c_3$, and $c_4$ are arbitrary parameters. Note here that unlike the case of (2+1) dimensional KdV equation only two arbitrary functions are allowed in (\ref{42}). 

Now substituting the above form (\ref{42}) for the manifold $\phi(x,y,t)$ into the truncated Painlev\'e series (\ref{38}) for the functions $u(x,y,t)$, $v(x,y,t)$ and $w(x,y,t)$, along with the expressions for the coefficient functions $u_0$, $v_0$, $w_0$, $u_1$, $v_1$, $w_1$, $v_2$ and $w_2$ given above, we finally obtain the solution to Eqs. (\ref{39}) as
\begin{subequations}
\begin{eqnarray}
u(x,y,t) &=& \frac{2 (c_2c_3-c_4c_1) f_x g_y}{[c_1+c_2 f(x,t)+c_3 g(y,t)+c_4 f(x,t)g(y,t)]^2}, 
\label{43_1} \\
      & &    (c_2c_3-c_4c_1) \neq 0,  \nonumber \\
v(x,y,t) &=& \frac{2 (c_2+c_4g)^2 f_x^2}{[c_1+c_2 f(x,t)+c_3 g(y,t)+c_4 f(x,t)g(y,t)]^2} \nonumber \\
& & -\frac{2 (c_2+c_4g) f_{xx}}{c_1+c_2 f(x,t)+c_3 g(y,t)+c_4 f(x,t)g(y,t)} \nonumber \\
& & +\frac{f_t+f_{xxx}+c f_x+(c_3+c_4 f)}{3a f_x},  \label{43_2} \\
w(x,y,t) &=& \frac{2 (c_3+c_4f)^2 g_y^2}{[c_1+c_2 f(x,t)+c_3 g(y,t)+c_4 f(x,t)g(y,t)]^2} \nonumber \\
& & -\frac{2 (c_3+c_4f) g_{yy}}{c_1+c_2 f(x,t)+c_3 g(y,t)+c_4 f(x,t)g(y,t)} \nonumber \\
& & +\frac{g_t+g_{yyy}+d g_y-(c_2+c_4 g)}{3b g_y}.  \label{43_3}
\end{eqnarray}  \label{43}
\end{subequations}
One can check that the above solutions of NNV equation reduces to the solutions of (2+1) dimensional KdV equation for the parametric choice $b=c=d=0$ and also by considering the truncated $w$ series to be zero.  One may note that the above form of solution coincides with the universal form of solutions reported by  Tang et al. \cite{ref5} and the special case $c_1=A$ ($A$ is arbitrary constant), $c_2=c_3=0$ was studied by Peng \cite{ref7_1}.  One can obtain periodic and localized dromion solutions here also following the procedure discussed in the earlier sections for the (2+1) dimensional KdV equation.

\section{Discussion}
In this paper, we have investigated the (2+1) dimensional K-dV equation and obtained
a four parameter solution involving three arbitrary functions  by using the Painlev\'e
truncation approach.  For different choice of parameters, we have generated two broad classes of localized coherent structures and elliptic function periodic wave solutions.  In particular, we have shown that the equation is trilinearizable. Even more general solutions can be generated from the trilinear equations which remains to be investigated.  
The existence of lower dimensional arbitrary functions helps us to construct novel localized solutions and study their interactions.   We have also extended the approach to generalized NNV equation and pointed out how similar solutions as that of the (2+1) dimensional KdV equation can be obtained.

\section{Acknowledgement}
The work of C. S. and M. L. form part of a Department of Science and
Technology, Govt. of India sponsored research project.  
R.R. wishes to thank the Department of Science and Technology (DST) for sponsoring a major research project.  The work of M. L. is supported by Department of Atomic Energy - Raja Ramanna Fellowship.
\\

\begin{appendix}
\section{One dromion solution through Hirota Bilinearization}
Here we briefly point out how the (1,1) dromion solution can be obtained through Hirota bilinearization
method \cite{ref4}.
To bilinearize equation (\ref{1}), we make the transformation
$$
u=-2 \partial_{xy}(\mbox{log}\phi), \;\;\; v=-2 \partial_{xx}(\mbox{log}\phi), \eqno{(A.1)}
$$
which can be identified from the Painlev\'e analysis.  The resultant bilinear form is
given by
$$
(D_yD_t+D_x^3D_y)\phi \cdot \phi = 0,  \eqno{(A.2b)}
\label{A1}
$$
where $D$'s are the usual Hirota operators.
To generate a (1,1) dromion, one considers the ansatz
$$
\phi = 1+e^{\psi_{1}}+e^{\psi_{2}}
+K e^{\psi_{1}+\psi_{2}}, \eqno{(A.3)}
$$
where
$$
\psi_{1} = k_1 x-k_1^3 t+\delta_1, \eqno{(A.4a)}
$$
$$
\psi_{2} = l_1 y+\delta_2. \eqno{(A.4b)}
$$
Here $k_1$, $l_1$, $\delta_1$, $\delta_2$ and $K$ are arbitrary constants. 
The (1,1) dromion is given by \cite{ref4}
$$
u = \frac{2 k_1 l_1 (1-K) e^{\psi_{1}+\psi_{2}}}{(1+e^{\psi_{1}}+e^{\psi_{2}}+K e^{\psi_{1}+\psi_{2}})^2}. \eqno{(A.5)}
$$
This is a special case of the dromion solution we have obtained in (\ref{14})
with the constants $\alpha=\beta=1$, $a=\frac{k_1}{2}$, $b=\frac{l_1}{2}$, $c=\frac{\pm 2}{\sqrt{1-K}}$, $c_1=-\frac{k_1^3}{2}$, $d_1=\frac{1}{2}\mbox{log}\frac{c-2}{c}+\frac{\delta_1}{2}$, $d_2=\frac{1}{2}\mbox{log}\frac{c-2}{c}+\frac{\delta_2}{2}$. 
Thus the standard dromion solutions become special cases of the general solutions obtained in section 3. 
\end{appendix}

{\bf Figure captions} \\
{\bf Figure 1} (a) Elliptic function solution (\ref{13}) for $q(x,y,t)$,(b) Localized dromion solution (\ref{14}) for $q(x,y,t)$ and (c) the solution corresponding to (\ref{15}) for $v(x,y,t)$ \\
{\bf Figure 2} (a) Elliptic function solution (\ref{17}) for $q(x,y,t)$, (b-d) (2,1) dromion solution (\ref{18}) for $q(x,y,t)$ and its interaction at time units
(b) $t=-4$, (c) $t=0$ and (d) $t=4$ \\
{\bf Figure 3} (a) Elliptic function solution (\ref{20}), (b) bounded two dromion solution (\ref{21}) \\
{\bf Figure 4} Singular solution for (\ref{14}) at different instants of time (a)$t$=-1, (b)$t$=0, (c)$t$=1  \\
{\bf Figure 5} (a) Elliptic function solution (\ref{24}) for $q(x,y,t)$
(b) Localized dromion solution (\ref{25}) for $q(x,y,t)$
(c) the solution corresponding to (\ref{26}) for $v(x,y,t)$ \\
{\bf Figure 6} (a) Elliptic function solution (\ref{28}) for $q(x,y,t)$, (b-d) (2,1) dromion solution (\ref{29}) for $q(x,y,t)$ and its interaction at time intervals
(b) $t=-4$, (c) $t=0$ and (d) $t=4$ \\
{\bf Figure 7} (a) Elliptic function solution (\ref{31}), (b) bounded two dromion solution (\ref{32}) \\


\begin{thebibliography}{15}
\bibitem{ref1}
Boiti M, Leon J J P, Martina L, Pempinelli F. Scattering of localized
solitons in the plane. Phys. Lett. A 1988; 132: 432-439. 
\bibitem{ref2} 
Fokas A S, Santini P M. Dromions and a boundary value
problem for the Davey-Stewartson I equation. Physica D 1990; 44: 99-130.
\bibitem{ref3} Ablowitz M J and Satsuma J. Solitons and rational solutions of nonlinear evolution equations. J. Math. Phys. 1978; 19: 2180-2186.
\bibitem{ref4} Radha R and Lakshmanan M. Singularity analysis and localized coherent structures in (2+1)-dimensional generalized Korteweg–de Vries equations. J. Math. Phys. 1994; 35: 4746-4756.
\bibitem{ref5} Tang X Y, Lou S Y, Zhang Y. Localized excitations in (2+1)-dimensional
systems. Phys. Rev. E.  2002; 66: 046601.
\bibitem{ref6} Konopelchenko B G. Solitons in multidimensions. Singapore: World Scientific; 
1993.
\bibitem{ref7} Radha R, Senthil Kumar C, Lakshmanan M, Tang X Y and Lou S Y. Periodic and localized solutions of the long wave-short wave resonance interaction equation. J. Phys. A: Math. Gen. 2005; 38: 9649-9663.
\bibitem{ref7_1} Peng Y. A class of doubly periodic wave solutions for the generalized Nizhnik-Novikov-Veselov equation. Phys. Lett. A. 2005; 337: 55-60.
\bibitem{ref7_2} Peng Y. Exact periodic and solitary waves and their interactions for the (2+1)-dimensional KdV equation. Phys. Lett. A. 2006; 351: 41-47.
\bibitem{ref8} Boiti M, Leon J J P, Manna M and Pempinelli F. On the spectral transform of a Korteweg-de Vries equation in two spatial dimensions. Inverse Problems. 1986; 2: 271-279.
\bibitem{ref9} Weiss J, Tabor M, and Carnevale G. The Painlev\'e property for partial differential equations. J. Math. Phys. 1983; 24: 522-526.
\end{thebibliography}
\end{document}